\documentclass[useAMS,usenatbib]{mn2e}
\input epsf.sty

\title[Time-delays in GRS 1915+105]{Time delays between the soft and hard X-ray
bands in GRS 1915+105}
\author[A. Janiuk \&  B. Czerny]{A. Janiuk$^{1}$\thanks{E-mail:
agnes@camk.edu.pl}, B. Czerny$^{1}$ 
\\
$^{1}$Nicolaus Copernicus Astronomical Centre, Bartycka 18,
            00-716 Warsaw, Poland}
\begin{document}


\pagerange{\pageref{firstpage}--\pageref{lastpage}} \pubyear{2004}

\maketitle

\label{firstpage}

\begin{abstract}
 The hard X-ray lightcurves exhibit delays of $\sim 1$ s with respect to the soft
X-ray lightcurves when the microquasar GRS 1915+105 is in the state of frequent, 
regular  outbursts (states $\rho$ and $\kappa$ of
Belloni et al. 2000). Such  outbursts are supposed to be driven by the radiation
pressure instability of the inner disc parts. The hard X-ray delays are then caused
by the time needed for the adjustment of the corona to changing conditions
in the underlying disc. We support this claim by the computation of the time
evolution of the disc, including a non-stationary evaporation 
of the disc and mass exchange with the corona. 
\end{abstract}


\section{Introduction}

The observations of Galactic black hole binaries (GBH) imply the coexistence
of a relatively cold, optically thick accretion disc, responsible
for a thermal disc-blackbody component in their soft X-ray spectra, 
with a hot, optically thin medium that is the source of power-law spectral tail 
in the hard X-ray band (see e.g. Done 2002 for a review). 
At least in the High and Very High spectral states, the latter may have the
 form of a corona above the accretion disc, which means that
the geometrical 
configuration of these two media is such that both are extending down to the last 
stable orbit around the black hole, being vertically separated from each other.
The hard X-rays are produced via Compton upscattering of seed
photons coming from the underlying disc, which basically requires the energy 
dissipation within the corona. Apart from this radiative coupling, also a 
mass transfer, i.e. evaporation and condensation of matter between the disc and corona
is possible (Meyer \& Meyer-Hofmeister 1994; R\'o\.za\'nska \& Czerny 2000).

Radiation pressure instability (e.g.
Taam \& Lin 1984; Lasota \& Pelat 1991)
 seems to be a plausible mechanism to
account for the characteristic variability of the Galactic microquasar GRS 1915+105
(Mirabel \& Rodriguez 1994; Taam, Chen \& Swank 1997;
Belloni et al. 2000). Exemplary lightcurves
were analyzed recently by Naik et al. (2002).
Some of these lightcurves exhibit a very regular shape of outbursts,
that can be possibly related to the disc variability, while others are more
chaotic and probably driven by other mechanism, e.g. jet emission.
Time-dependent accretion disc model with jet emission 
was studied e.g. by Nayakshin et al. (2000) and Janiuk et al. (2002).
Recently, Watarai \& Mineshige (2003)
analyzed the oscillations of this source allowing for the temporary evacuation of
the inner disc.

If the disc instability is the
primary cause of regular outbursts, we may expect that the hard X-ray coronal emission 
would be significantly delayed with respect to the soft X-ray disc emission.
In the present paper we check the hypothesis of the radiation pressure
instability as the outburst driver by analyzing the observed time delays
between the soft and hard X-ray emission. We study the behaviour
of GRS 1915+105 in various variability classes.
We compare the observed time delays with the exemplary results of our
theoretical model. In this model we compute numerically
the time
evolution of the disc-corona system with the mass exchange between the two media.

The structure of this article is as follows. 
In Section \ref{sec:delays} we present the results of analysis
of the microquasar GRS 1915+106 observations, obtained by the {\it Rossi X-ray 
Timing Explorer}
(RXTE). The X-ray lightcurves
were examined to determine the time lags between hard and soft X-ray bands
in different modes characteristic for the variability of this source.
In Section \ref{sec:method} we
describe our model and assumptions about the disc and corona structure. The 
initial state of the disc plus corona system,
as well as the prescription for the mass exchange, are described 
in Section \ref{sec:assum}, 
while the time-dependent equations, according to which this system subsequently 
evolves with time, are given in Section \ref{sec:evol}.
The results of the evolution are given in Section \ref{sec:results}. We discuss our 
model and compare its predictions with observations in Section \ref{sec:diss}.
The conclusions are given in Section \ref{sec:con}.

\section{Time delays}
\label{sec:delays}

Fourier resolved time delays in the lightcurves of GRS 1915+105 were analysed
by Muno et al. (2001) for the periods of extended hard steady
state. In the present paper we concentrate mostly on periods with
significant outbursts. In this case Fourier resolved phase lag approach is
not best suited since clear semi-periodic signal dominates each lightcurve.
Therefore, we restore to the simplest direct delays, as measured by the
cross-correlation function.


In order to determine the time lag between soft and hard X-ray emission we
performed the Fourier analysis of these lightcurves by means of the Fast Fourier 
Transform (FFT) method. The cross-correlation function of two periodic functions 
$F(t)$ and $G(t)$ is defined as:
\begin{equation}
Corr (\Delta t) = \int F(t) G(t + \Delta t) dt.
\label{eq:crossfun}
\end{equation}

\begin{figure}
\epsfxsize = 250pt
\epsfbox{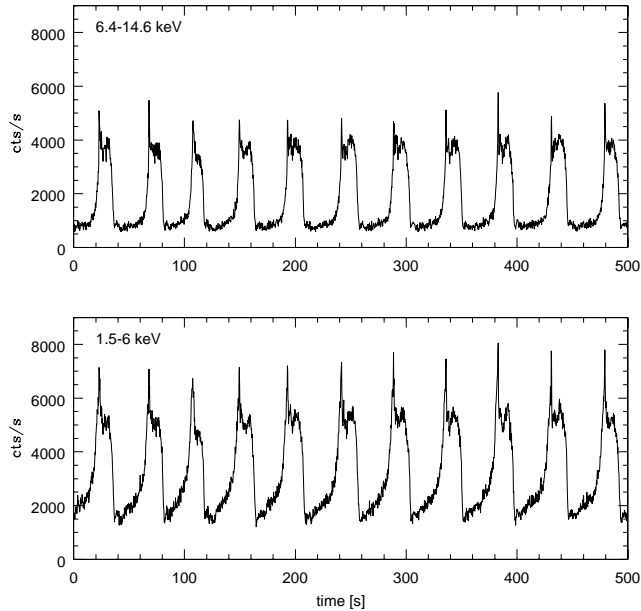}
\caption{The X-ray lightcurves of GRS 1915+105 obtained from RXTE observation
from June 20 2000 (class $\rho$), in the energy bands 1.5-6 keV and 6.4 - 14.6 keV.
\label{fig:obs}}
\end{figure}

For our analysis we select lightcurves representative for various variability
classes, as studied in detail in Belloni et al. (2000).
We choose the exemplary observations of GRS 1915+105 made between 1996 and 2000,
available through the public RXTE archive.
The data were binned to 0.256 seconds and the lightcurves were generated for the
two energy bands separately: 1.5-6 keV (PHA channels 0-14) and 6.4-14.6 keV 
(PHA channels 15-35). Each lightcurve consists of one or more intervals,
and in the subsequent analysis we compare the single intervals
between each other.

\begin{figure}
\epsfxsize = 250pt
\epsfbox{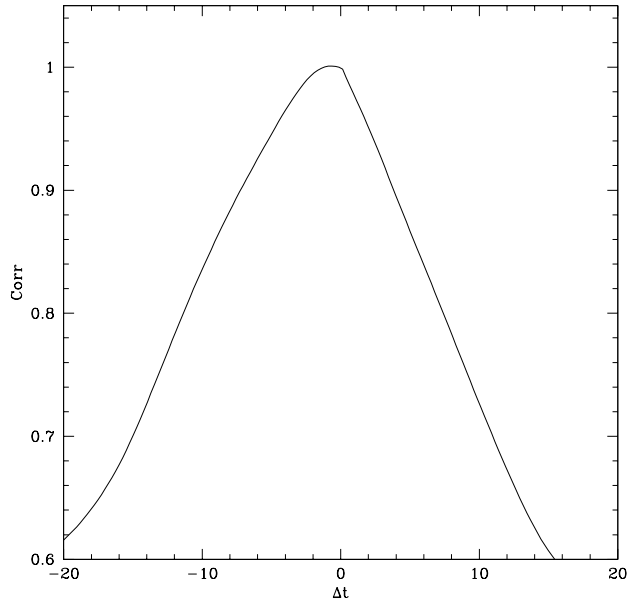}
\caption{The normalized cross-correlation function of the two lightcurves,
 1.5-6 keV and 6.4 - 14.6 keV, shown in Fig. \ref{fig:obs}. The main peak is
shifted to -0.768 s, which means that the hard X-ray lightcurve lags the soft X-ray 
one by 0.768 s.
\label{fig:corr}}
\end{figure}

In Figure \ref{fig:obs} we show an exemplary lightcurve of the microquasar 
observed by RXTE on 20 June, 2000. 
This variability pattern belongs to the  class $\rho$ of Belloni et al. (2000).
The cross-correlation function
is calculated for the two lightcurves obtained in this observation
and shown in Figure \ref{fig:corr}.
The maximum of this cross-correlation function defines the time lag, $\Delta t$,
between the two lightcurves. 
In this case it is equal to $\Delta t =  0.768$ s. This means that the hard X-ray 
lightcurve lags the soft X-ray one by 0.768 seconds, which in this case is 
roughly 1.5\% of the outburst duration.
Other results are given in Table \ref{tab:tab1}.

\begin{table}
\caption{Time lag of the hard X-ray lightcurve (6.6-14.6 keV) with respect to the 
soft X-ray lightcurve (1.5-6 keV). $\Delta t_{1}$ is calculated for the non-smoothened 
lightcurves, binned by 0.256 sec., and $\Delta t_{2}$ is calculated for the  
lightcurves smoothened over $\Delta T =1$ s.
}
\label{tab:tab1}
\begin{tabular}{c|c|c|c|c}
\hline
PID  & Date obs.  & Class &  $\Delta t_{1}$ [s] &  $\Delta t_{2}$ [s] \\
\hline
20402-01-28-00 & 18/05/97 & $\alpha$ & 0.0 & 0.0 \\
20402-01-41-00 & 19/08/97 & $\delta$ & 0.0 & 0.0 \\
20402-01-37-00 & 17/07/97 & $\gamma$ & 0.0 & 0.0 \\
20402-01-44-00 & 31/08/97 & $\beta$ & 0.0 & 0.0 \\
10408-01-15-00 & 16/06/96 & $\theta$ & 0.0 & 0.0 \\
20402-01-36-00 & 10/07/97 & $\lambda$ & 0.0 & 0.256 \\
20402-01-33-00 & 18/06/97 & $\kappa$ & 0.256 & 0.512 \\
50703-01-15-01 & 20/06/00 & $\rho$    & 0.768 & 1.024 \\
\hline
\end{tabular}
\end{table}

 Regular, large timescale outbursts are characteristic for the 
lightcurves of type $\rho$ and $\kappa$.
In other cases,
the rapid, stochastic variability is overimposed on the more regular
outburst pattern of a longer timescale (classes $\lambda$, $\beta$, and $\theta$)
or even dominates the whole observation interval (classes $\alpha$, 
$\delta$ and $\gamma$). This rapid variability 
does not exhibit any time lags between hard and soft X-ray bands and is 
probably not correlated with the disc/corona system instability.
Therefore, in order to determine more accurately the correlation between hard and soft 
X-rays in case of long outbursts, we filter out the rapid, small scale variations.
We smoothen the lightcurves by applying the running mean filter
(Stull 1988), to remove the variations
shorter than (arbitrary) $\Delta T=1$ sec. The resulting time lags are given
in the last column of Table \ref{tab:tab1}.
We see that the time lags between regular outbursts are even more pronounced,
if these outbursts are cleared from the stochastic variability.
On the other hand, the time lag $\Delta t_{2}$ 
remains only marginally detectable, or equal to zero,
in the lightcurves that 
exhibit mostly the stochastic variations.

The measured  lags calculated for the two latter observations, $\kappa$ and $\rho$, seem to
 be exceptionally high
for an X-ray binary (\.Zycki, private communication). However, this is not
surprising since only GRS~1915+105 exhibits outbursts while other sources show
only stochastic type of variability, which must be of different nature.

\section[]{Time-dependent disc/corona model}
\label{sec:method}

Accretion discs are known to be locally unstable in certain temperature
(corresponding to the accretion rate)  and density ranges, in which 
the cooling and heating balance is strongly influenced
either by atomic opacities or by the radiation pressure. Such instabilities
do not disrupt the disc completely, but
lead to repetitive outbursts.

The first type of instability, connected with the partial hydrogen ionization
(Smak 1984; Meyer\& Meyer-Hofmeister 1984), 
operates in accretion discs of binary systems in the range of radii of the 
order of $10^{4} - 10^{5} R_{\rm Schw}$ and is responsible for the luminosity
changes on timescales of months, e.g. Dwarf Nova outbursts.

The other instability, caused by the radiation pressure domination over the gas 
pressure, operates in the innermost regions of the disc, where the temperature exceeds
$\sim 10^{6}$ K, and is responsible for the disc variability on the shortest timescales
(of the order of tens - hundreds of seconds).
This radiation pressure
instability was first noticed in Pringle, Rees \& Pacholczyk (1974) and
studied in Lightman \& Eardley (1974) and Shakura \& Sunyaev (1976).
Abramowicz et al. (1988) found that radial advection has
stabilizing effect on the disc at high accretion rates and the
time-dependent calculations of the disc limit-cycle behaviour 
were performed by Honma et al. (1991) 
and by Szuszkiewicz \& Miller (1998).


The innermost disc regions in which the radiation pressure instability is
possible, are covered by the hot corona. Direct comparison of the extension
of the radiation pressure domination zone and the corona covered zone, resulting from 
the model of the stationary, two-temperature corona, was performed by 
Janiuk \& Czerny (2000). In the time-dependent model of radiation-pressure 
instability proposed in Janiuk, Czerny \& Siemiginowska (2002) we used a simplified
description of a stationary corona above the fluctuating disc,
parameterized by a constant value of the fraction of gravitational energy
dissipated in the corona ($f_{\rm cor}$). Here we extend our model to the case
of a non-stationary corona, that forms due to the continuous evaporation of material
form the disc surface. 
Due to the mass exchange with the disc the corona follows its time-dependent behaviour
and therefore periodic changes are expected also in the hard X-ray luminosity.

\subsection{Assumptions and model parameters}
\label{sec:assum}

\subsubsection{Disc}

First we describe the initial steady state disc model, from which we
start our subsequent calculations.
Throughout the calculations we use the vertically integrated equations
of the disc structure, as 
the disc geometrical thickness $H$ is always small 
 ($H/r \sim 0.01$ in the quiescent disc and  
$H/r \sim 0.1$ in the outburst; see Section \ref{sec:stabil}).

The angular velocity of the disc is assumed to be Keplerian,
$\Omega = \sqrt{GM/r^{3}}$, and the sound speed is $c_{\rm
s}=\sqrt{P/\rho} = \Omega H$. 
Here $\rho$ is the gas density in g/cm$^{3}$, $M$ is the mass of the 
accreting black hole and $G$
 is the gravitational constant.
A non-rotating, Schwarzschild black
hole is assumed and the inner radius of the disc is always at 3
$R_{\rm Schw}$. The outer radius is equal to 300 $R_{\rm Schw}$, and at this radius
a constant mass inflow, parameterized by the external 
accretion rate $\dot M_{\rm ext}$ 
is assumed. Only the innermost zone up to  $\sim 100 R_{\rm Schw}$ is the subject
to radiation pressure instability, while the rest of the disc is stable
(the exact value of the radial extension of the unstable zone 
depends on the model; see Section \ref{sec:stabil}).
The mass of the black hole is assumed to be $10 M_{\odot}$.

For the disc heating we 
assume that the viscous stress 
tensor is proportional to the total pressure $P$:
\begin{equation}
\tau_{r\varphi}=-\alpha P,
\end{equation}
and the vertically integrated heating rate is
\begin{equation}
Q^{+}_{\rm visc} = {3 \over 2}\alpha\Omega H P
\label{eq:heat}
\end{equation}
 where $\alpha$ is the viscosity parameter given by Equation 
\ref{eq:visclaw}.
The total pressure $P$ consists of the gas and radiation pressure:
\begin{equation}
P = P_{\rm rad} + P_{\rm gas}
\label{eq:ptot}
\end{equation}
\begin{equation}
P_{\rm gas} = {k \over m_{\rm p}}\rho T
\end{equation}
\begin{equation}
P_{\rm rad} = {1 \over 3} a T^{4}
\end{equation}
where  $T$ is the 
mid-plane temperature, and $k$, $m_{p}$ and $a$ are physical constants.

The angular momentum transport in the accretion disc is driven mostly by
the magneto-rotational turbulences (see Janiuk et al. 2004 and references therein).
Since the magnetic fields are quickly expelled from the radiation-dominated disc 
(Sakimoto \& Coroniti 1989), the transport efficiency, and in turn 
the viscosity parameter, must decrease when the radiation pressure becomes dominant.
On the other hand, when the contribution from radiation pressure is only moderate,
it may still be possible to couple the radiation to the particles.
Here we adopt a modified viscosity law for the accretion disc 
(Chen \& Taam 1993; Nayakshin, Rappaport \& Melia 2000):
\begin{equation}
 \alpha =\alpha_{0}  {(1+\xi/\xi_{0}) \over (1+(\xi/\xi_{0})^{2} }
\label{eq:visclaw}
\end{equation}
where $\xi = P_{\rm rad}/P_{\rm gas}$ (for the discussion of this
parameterization see Section \ref{sec:viscpar}). For the model parameters 
$\alpha_{0}$ and $\xi_{0}$
we assumed the values 0.01 and 8.0, respectively.
This prescription implies that for small to moderate values of $\xi$ 
we have effectively the disc heating proportional to the total pressure,
while for large values of $\xi$ the viscosity is proportional to the gas pressure.
Therefore the radiation pressure instability may still operate, 
contrary to the so-called $\beta$-disc prescription (Lightman \& Eardley 1974).

The cooling in the disc is due to advection and radiation and 
the radiative cooling is equal to:
\begin{equation}
Q^{-}_{\rm rad}={ P_{\rm rad} c \over \tau}={ \sigma T^{4} \over 
\kappa \Sigma}
\label{eq:qrad}
\end{equation}
where $\tau$ is the optical depth, $\Sigma = \rho H$ is the gas column density
in g cm$^{-2}$, $c$ and $\sigma$ are physical constants,
and we adopt the electron scattering opacity
$\kappa=0.34$ cm$^{2}$/g.

The advective cooling in a stationary disc is determined from the global ratio 
of the total
advected flux to the total viscously generated flux (e.g. Paczy\' nski \&
Bisnovatyi-Kogan 1981; Muchotrzeb \& Paczy\' nski 1982; Abramowicz et al. 1988)
\begin{equation}
Q^{-}_{\rm adv}= {F_{\rm adv} \over F_{\rm tot}}=
- {2 r P q_{\rm adv} \over 3 \rho GM}
\label{eq:fadv}
\end{equation}
and
\begin{equation}
q_{\rm adv}=(12-10.5 \beta) {\partial \ln T \over \partial \ln r} - 
(4-3 \beta){\partial \ln \rho \over \partial \ln r}
\label{eq:qadv}
\end{equation}
Here $\beta$ is the ratio of the gas pressure 
to the total pressure
$\beta = P_{\rm gas}/P = 1/(1+\xi)$. 
In the initial stationary disc we  assume 
that $q_{\rm adv}$ is approximately constant and of the order of unity
(in the subsequent evolution the advection will be calculated more
carefully, with appropriate radial derivatives).
 
In order to calculate the initial steady-state configuration, we solve the
energy balance: $F_{\rm tot} = Q^{+}_{\rm visc} = Q^{-}_{\rm
adv}+Q^{-}_{\rm rad}$.
Here the total energy flux $F_{\rm tot}$ dissipated within the disc at a 
radius $r$ is calculated as:
\begin{equation}
F_{\rm tot} = {3 G M \dot M \over 8 \pi r^3} f(r)
\label{eq:ftot}
\end{equation}
where
\begin{equation}
f(r) = \Big(1-({3 R_{\rm Schw}\over r})^{3/2}\Big){r-R_{\rm Schw}\over 2 R_{\rm Schw}}
\end{equation}
is the boundary condition in the pseudo-Newtonian potential 
(Paczy\'nski \& Wiita 1980).
We choose the initial accretion rate {\bf $\dot M$} that is constant throughout the disc and is 
low enough for the disc to be on the stable gas-pressure dominated
branch, where neither radiation pressure nor advection is important (see Figure
\ref{fig:scurves}).
The initial model is calculated by means of a simple Newtonian method, 
through which we determine the radial profiles of
density and temperature, as well as the disc thickness.

\subsubsection{Corona}

The corona is assumed to be geometrically thick and optically thin. Therefore its
height is assumed to be equal to radius, $H_{\rm cor}=r$, and the pressure in the corona
is only gas pressure due to ions:
\begin{equation}
 P_{\rm cor} =  P_{\rm gas} = {k \over m_{\rm p}} \rho_{\rm cor} T_{\rm cor}
\end{equation}
The contribution from the electrons is neglected. The corona is hot and its ion 
temperature is assumed to be equal to the virial temperature:
\begin{equation}
T_{\rm cor} = T_{\rm vir} = {G M \over r}{m_{\rm p} \over k}
\end{equation}

The initial configuration of the corona is computed under the assumption that 
its optical depth is equal to unity: 
\begin{equation}
\tau_{\rm cor}  = \kappa \Sigma_{cor} = 1.0
\label{eq:taucor}
\end{equation}
and therefore the corona has a uniform surface density. When the time evolution starts,
the proper solution develops in the middle parts 
of the disc, but will be fixed at the boundaries $R_{\rm in}$ and $R_{\rm out}$ 
by the above condition.

\subsubsection{Mass exchange  (prescription I)}

The mass exchange rate in the vertical direction, between the disc and corona,
is equal to the ratio between the locally generated flux used to evaporate the
 disc material and the energy change per particle:
\begin{equation}
\dot m_{\rm z} = {F \over \Delta E/m_{\rm p}} = {F m_{\rm p} \over k T_{\rm cor}} 
\end{equation}
(measured per surface unit, g/s/cm$^{2}$). 

In a stationary disc the generated flux depends on the accretion rate and the
disc radius (see Eq.~\ref{eq:ftot}). We assume that during the time evolution
the energy flux leading to evaporation preserves this dependence. 
Since both the energy dissipated within the
corona and within the disc  can lead to disc evaporation,
we assume that the energy flux is proportional to the
sum of the disc and corona accretion rates, taken with different numerical 
coefficients:
$F \propto (0.5 B_{1} \dot M_{\rm cor} + B_{2} \dot M_{\rm disc})$. 
The coefficients 
$B_{1}$ and $B_{2}$ are in the range from 0 to 1 and express the fraction of
the generated flux that is used to drive the evaporation. The share of the corona
is  always lower than  the half of the total corona flux, since half of the flux from 
the corona is directed toward the observer, whereas the other half is directed towards the
disc and there reprocessed.

The total accretion rate is $\dot M = \dot M_{\rm cor} + \dot M_{\rm disc}$
and may be locally constant in case of a stable disc. When the disc is unstable,
both local accretion rates strongly depend on time and radius, and the
relative contribution of the disc and corona to the total flow also vary.

Expressing the accretion rates through the local variables 
$\dot M_{\rm cor} = 2\pi\Sigma_{\rm cor} r v_{\rm r}^{\rm cor}$ and 
$\dot M_{\rm disc} = 2\pi\Sigma_{\rm disc} r v_{\rm r}^{\rm disc}$ in the corona and disc 
respectively, we obtain a useful formula for the vertical mass transfer:
\begin{equation}
\dot m_{\rm z} = {3 \over 4 r} f(r)
         (0.5 B_{1} \Sigma_{\rm cor} v_{\rm r}^{\rm cor} + B_{2} \Sigma_{\rm disc} 
v_{\rm r}^{\rm disc})
\label{eq:dotmz}
\end{equation}
This formula is qualitatively similar to $\dot m_{\rm z} \sim \dot M_{\rm cor}^{5/3}
/r^{3/2}$, derived by R\'o\.za\'nska \& Czerny (2000) under specific assumptions about 
the disc/corona coupling. 
However, their model described a stationary evaporation and did not depend on time.
In our case the corona has to develop above the disc from the initial, 
uniform, very low density distribution, and this rise is provided by
the sum of disc and corona accretion rates. 
(In the above, we denoted the quantities in the disc with the subscript ``disc'', in 
order to discriminate them from the coronal ones. 
Please note, that whenever the physical 
quantities appear without any subscript, they also refer to the disc, and the coronal 
quantities are always distinguished with the subscript ``cor''.)

\subsection{Time evolution}
\label{sec:evol}

Having computed the initial disc and corona state we allow the density and
temperature of the disc and the density of the  corona to evolve with time. 
We solve the equation of mass and
angular momentum conservation:
\begin{equation}
{\partial \Sigma \over \partial t}={1 \over r}{\partial \over \partial
r}(3 r^{1/2} {\partial \over \partial r}(r^{1/2} \nu \Sigma))
- \dot m_{\rm z}
\end{equation}
and the energy equation:
\begin{eqnarray}
{\partial T \over \partial t} + v_{\rm r}{\partial T \over \partial r}
= {T \over \Sigma}{4-3\beta \over 12-10.5\beta}
({\partial \Sigma \over
\partial t}+  v_{\rm r}{\partial \Sigma \over \partial r}) \\
\nonumber +{T\over P H}{1\over 12-10.5\beta} 
(Q^{+}-Q^{-}).
\end{eqnarray}
Here  
\begin{equation}
v_{\rm r} = {3 \over \Sigma r^{1/2}} {\partial \over \partial r}
(\nu \Sigma r^{1/2})
\end{equation}
is the radial velocity in the disc while
$\nu=(2P\alpha)/(3\rho\Omega)$ 
is the kinematic viscosity. 
The heating term is given by Equation \ref{eq:heat}
and the  cooling term $Q^{-}$ is given by Equation \ref{eq:qrad}, 
while the advection is included in the energy equation
via the radial derivatives.

The evolution of the coronal density is given by mass and angular momentum conservation in the corona:
\begin{equation}
{\partial \Sigma_{\rm cor} \over \partial t}={1 \over r}{\partial \over \partial
r}(3 r^{1/2} {\partial \over \partial r}(r^{1/2} \nu_{\rm cor} \Sigma_{\rm cor}))
+ \dot m_{\rm z}.
\end{equation}
The radial velocity in the corona is calculated as:
\begin{equation}
v_{\rm r}^{\rm cor} = {3 \over \Sigma_{\rm cor} r^{1/2}} {\partial \over \partial r}
(\nu_{\rm cor} \Sigma_{\rm cor} r^{1/2})
\end{equation}
with $\nu_{\rm cor}=(2P_{\rm cor}\alpha_{\rm cor})/(3\rho_{\rm cor}\Omega)$ and
constant viscosity parameter in the corona $\alpha_{\rm cor}=0.01$.
There is no need to consider the thermal evolution of the corona, since its 
temperature is always equal to the virial temperature and does not vary with time.

We solve the above set of three time-dependent equations using the convenient
change of variables, $y=2r^{1/2}$ and $\Xi = y \Sigma$, at the fixed
radial grid, equally spaced in $y$ (see Janiuk et al. 2002 and references therein). 
The number of radial zones is set
to 216. After determining the solutions for the first 600 time steps by the
fourth-order Runge-Kutta method, we use the Adams-Moulton 
predictor-corrector method, allowing the time-step to vary,
when needed. 

We choose the no-torque  inner boundary condition, $\Sigma_{\rm in} =
T_{\rm in} = 0$ for the disc. The outer boundary of the disc is parameterized by an
external accretion rate $\dot M_{\rm ext}$. If this 
accretion rate is high enough, the inner disc parts gradually heat
themselves and finally end in the unstable regime, forcing the disc to
oscillate.
The boundary conditions in the corona are given by Equation \ref{eq:taucor}.

\section{Results}
\label{sec:results}

\subsection{Surface density and temperature evolution}
\label{sec:stabil}

 The local solutions of the accretion disc model, 
in the surface density vs. temperature ($\Sigma$- $T$) plane, can be calculated
for a stationary disc in the range of accretion rates.
(Alternatively, on the vertical axis we can have accretion rate $\dot M$ instead of
the disc temperature.) These solutions lie along the S-shaped stability curve,
whose position on the diagram depends on the model parameters: black hole mass,
viscosity and radius (c.f. Janiuk et al. 2002, Figs. 1, 3).

 Both upper and lower branches 
of the {\bf S}-curve are viscously and thermally 
stable. 
On the lower stable branch the gas pressure dominates; 
the middle branch is unstable (radiatively cooled and
radiation pressure dominant), as shown in detail by 
Pringle, Rees \& Pacholczyk (1974) and Lightman \& Eardley (1974). 
The upper branch is stabilized in our model again by the dominant gas pressure,
due to the modified viscosity law. In case of the standard viscosity 
(with constant $\alpha$) this branch would be stabilized mainly by advection, as
shown in Abramowicz et al. (1988). In our case the advection is also taken into 
account, but its role is never dominant.

The {\bf S}-curve can also be plotted in the $\dot M - \Sigma$ plane,
for any chosen disc radius. This means that the temporary local solutions are
determined by
the mean (i.e. external) accretion rate
in the disc. Whenever the external accretion rate is low, so that at
all the radii in the disc the local solution sits on the lower, stable
branch, the accretion proceeds with this rate, 
which is constant throughout the disc and constant in time.
But if the accretion rate is higher than some critical value, $\dot M_{\rm ext} > 
\dot M_{\rm crit}$,
the solutions in the innermost annuli will find themselves on the unstable branch.
The higher $\dot M_{\rm ext}$, the more disc annuli will be unstable.
This leads to the disc fluctuations, since the accretion cannot proceed smoothly
in the unstable mode. Therefore the local accretion rate 
in the innermost strips changes periodically
between the lower and upper stable solutions, being no longer equal to 
$\dot M_{\rm ext}$ (the accretion rate starts to depend on radius and time). 
This is displayed in the local diagrams $\Sigma -T$
that are resulting form the {\it time-dependent} model.

\begin{figure}
\epsfxsize = 250pt
\epsfbox{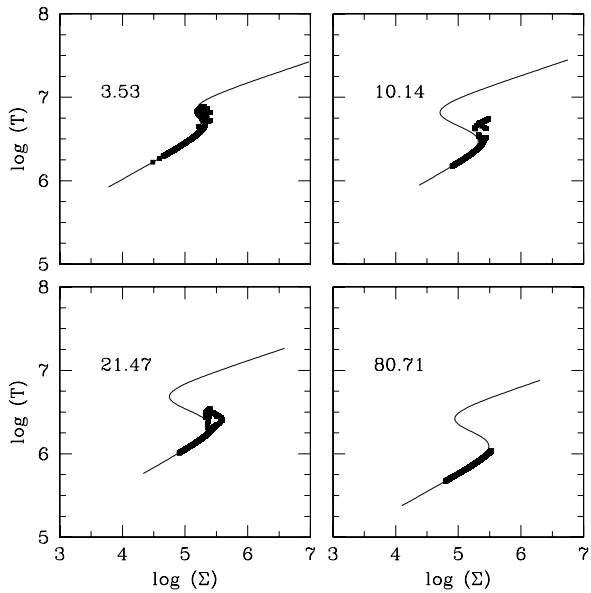}
\caption{Local evolution of the disc on the surface density -
temperature plane, plotted for 4 values of radius: 3.53, 10.14, 21.47
and 80.71 $R_{\rm Schw}$. 
The thin solid line  marks the stability curve resulting from the initial
steady disc model, and the solid points are the 
time-dependent solutions.
\label{fig:scurves}}
\end{figure}

The exemplary  stability curves of the accretion disc, 
calculated at several radii from the stationary disc model,
are shown in Figure \ref{fig:scurves} (thin solid lines).
The thick points represent the subsequent solutions of the time-dependent model.

The evolution of the disc on the surface density - temperature plane proceeds 
at first along the lower stable branch, up to the instability region.
This is forced by the value of the external accretion rate parameter,
which has to be large enough to drive the disc to the unstable configuration.
Here we assumed $\dot M_{\rm ext} = 1.5 \times 10^{19}$ g s$^{-1}$,
which is equal to 0.45 of the Eddington rate (for black hole mass of 10 $M_{\odot}$
and efficiency of 1/16).
The critical accretion rate 
in our model depends on whether the corona covers the disc or not;
$\dot m_{\rm crit} = 0.05$ for the plain disc, while in the case of a 
disc/corona system the corona has a stabilizing role and the critical accretion rate
is about $\dot m_{\rm crit} = 0.2$.
The accretion rates 
within the range $7.5 \times 10^{18}- 2.6 \times 10^{19}$ g s$^{-1}$,
were obtained for soft states of GRS 1915+105 by Sobolewska \& \. Zycki (2003).

Firstly, in the starting, steady configuration we assumed the accretion rate of
$\dot m \sim 1.5\times 10^{-2}$ of the Eddington rate throughout the disc. 
Therefore at the beginning of the subsequent time-dependent calculations
the model has to saturate at the temperatures and densities imposed
by the value of $\dot M_{\rm ext}$, imposed at the outer disc radius.

Next, the evolution proceeds 
in a form of loops between the lower and upper branches.
Each loop refers to a single cycle of the instability, and the size of this loop
depends on the location in the disc. For larger radii the loops become smaller
and finally, in the outer disc regions (above $\sim 80-100  R_{\rm Schw}$), 
there are no instabilities
and the disc remains stable all the time.
The exact value of the maximal radius of the instability zone depends
again on the model. In case of a plain disc,
for the external accretion rate  $\dot m_{\rm ext}=0.45$ it is 
$R_{\rm max} = 100 R_{\rm Schw}$, while for $\dot m_{\rm ext}= 0.56 $ it is 
$R_{\rm max} = 110 R_{\rm Schw} $. 
In case of the disc with corona the extension of the unstable zone is
respectively $R_{\rm max} = 80 R_{\rm Schw}$ and $R_{\rm max} = 90 R_{\rm Schw}$.

Note, that in the initial steady disc model we use a simplified parameterization
of advection, with $q_{\rm adv}=1.0$ (see Equation \ref{eq:qadv}). Therefore
the upper stable branch does not represent exactly the advective branch that
results from the time-dependent calculations, which are based on the
equations with radial derivatives. In fact, $q_{\rm adv}$ is not constant 
throughout the disc and should depend on radius. 
However, this initial simplification does not influence our results, since
our starting model is located on the lower, gas pressure dominated branch.
Here the advection is negligible, and the subsequent time-dependent
solutions match the stability curve perfectly.

\begin{figure}
\epsfxsize = 250pt
\epsfbox{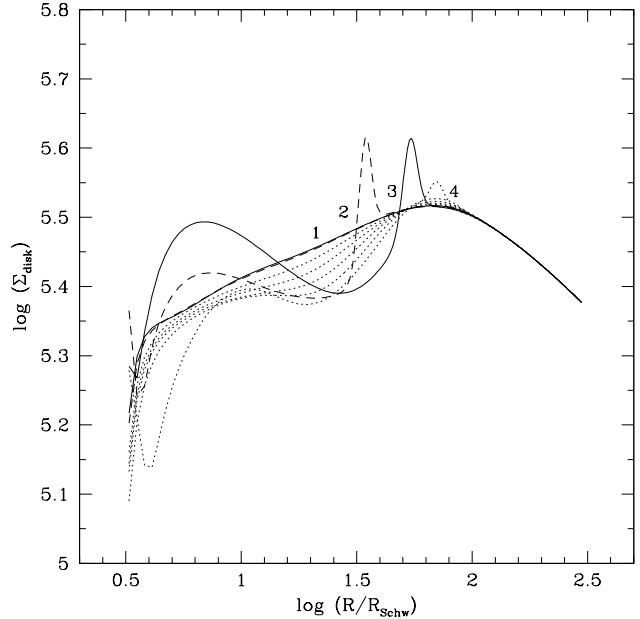}
\caption{Radial profiles of the disc surface density during the cycle of the 
evolution. The snapshots were made every 80 seconds and the whole cycle lasts
for 625 seconds. 
he solid curve (1) refers to the minimum of 
the cycle (lowest disc luminosity,
see also Figs. \ref{fig:sigcorev} and \ref{fig:lcurve}), 
the dashed curve (2) refers to the 
rise phase and the solid curve (3) refers to the maximum of the cycle. The dotted
line (4) and other dotted lines (unnumbered) refer to the decay phase.
\label{fig:sigev}}
\end{figure}

In Figure \ref{fig:sigev} we show the radial profiles of the surface density 
in the accretion disc,
in several snapshots during such a loop (one instability cycle). In the minimum of
a cycle the surface density in the inner parts of the accretion disc has a flat
radial distribution. When the outburst starts, there appears a sharp density 
peak in the outer part of the unstable zone, which then moves outward.
In the maximum of the cycle this peak is accompanied by the largest fluctuation in the
density distribution. 
The reason for these fluctuations are
 the viscosity and angular momentum transfer changes
inside this propagating ``density wave''.
When the innermost radii of the disc switch to the hot state,
the geometrical thickness of this region also rises, thus giving the rise to the
kinematic viscosity. The increased transport rate results in the
temporary density decay in the unstable zone, as the material starts to fall faster 
into the black hole. Simultaneously, at the inner edge of the disc there forms
a temporary ``bump'' of material, forced by the no-torque inner
boundary condition.

The density fluctuation subsequently vanishes in the end
of the cycle and during the decay phase the 
surface density in the unstable zone gradually rises, to reach the starting
configuration.

The corona evolves on a timescale much shorter than the disc. 
First, we investigate the
corona formation in case of no mass exchange with the disc ($B_{1}=B_{2}=0$).
The initial distribution of the surface density in the corona was flat, as 
determined by Equation \ref{eq:taucor}. When the evolution starts,
the corona very quickly achieves its final shape -- the surface density distribution 
saturates after $\sim$125 sec. At the same time the disc evolves very slowly,
being ready to start its oscillations after $\sim 10^{4}$ sec. of a gradual rise
in density and temperature.
Since there is no coupling with the corona, these oscillations do not
influence the corona structure. 

\begin{figure}
\epsfxsize = 250pt
\epsfbox{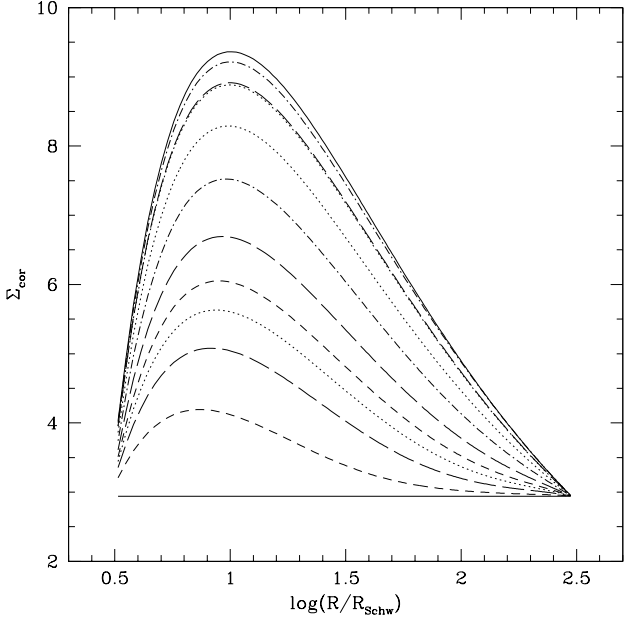}
\caption{Radial profiles of the corona surface density
in case of no mass exchange with the disc ($B_{1}=B_{2}=0$). The profiles were 
calculated every 10 seconds and the density saturated at its final profile
after $\sim 125$ seconds of the evolution. 
\label{fig:sigcor}}
\end{figure}

Secondly, we proceed with the evolution in case of a substantial mass exchange 
between the disc and corona ($B_{1}=B_{2}=0.5$).
In this case, the evaporation of the disc accelerates the corona formation,
and after $\le$ 100 seconds the coronal surface density in the maximum
of the radial distribution (around $10 R_{\rm Schw}$) exceeds 10 g/cm$^2$.
Then the corona is further fed with material by the disc as its evolution proceeds
along the stability curve and the coronal surface density further gradually increases.
The maximum surface density saturates at  $\sim 80$ g/cm$^2$ after 
the disc reaches the critical point on the stability curve, which determines
the maximum density in the disc.

\begin{figure}
\epsfxsize = 250pt
\epsfbox{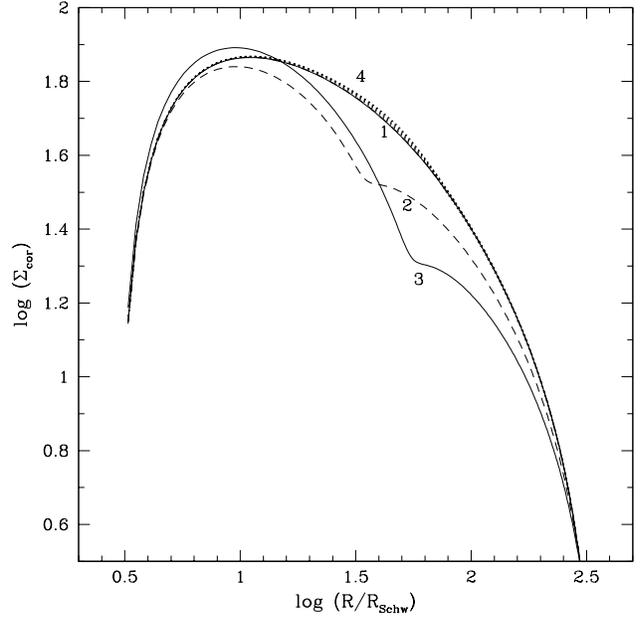}
\caption{Radial profiles of the corona surface density during the cycle of the 
evolution. The snapshots were made every 80 seconds and the whole cycle lasts
for 625 seconds. The labels are the same as in Fig. \ref{fig:sigev}.
\label{fig:sigcorev}}
\end{figure}

When the disc oscillations start, the corona also follows its time-dependent behavior.
In Figure \ref{fig:sigcorev} we plot the surface density distribution in the corona
during one cycle of the disc instability. The curves are plotted every 80 seconds
in case of a full cycle lasting $\sim 625$ seconds. The solid line (1) 
is the coronal density in the minimum of the disc limit cycle,
the  dashed line (2) corresponds to the rise phase of the disc outburst and the  
solid line (3) corresponds to the maximum of the cycle (outburst). Immediately
after the outburst the coronal density distribution comes back to the initial
configuration (dotted lines) and remains there for the most of the cycle (4).
Therefore during the decay phase of the disc, when the disc surface density gradually 
changes to reach a flat distribution, there is no change of the density
in the corona.

\begin{figure}
\epsfxsize = 250pt
\epsfbox{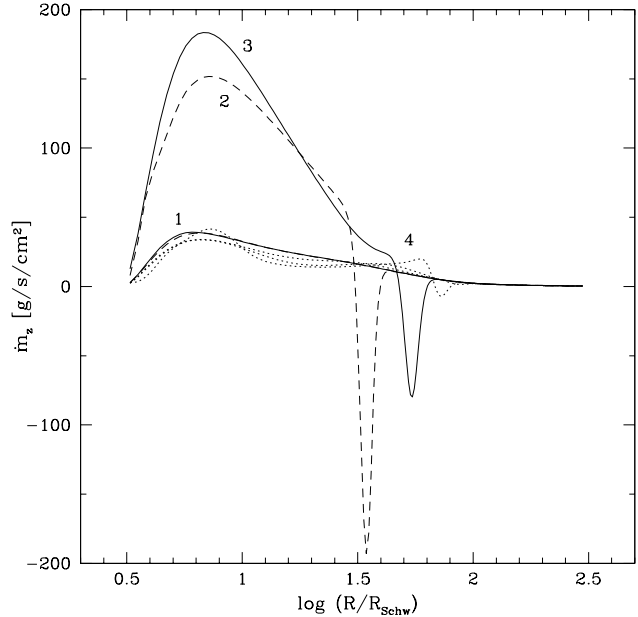}
\caption{Radial profiles of the mass exchange rate during the cycle of the 
evolution. The snapshots were made every 80 seconds and the whole cycle lasts
for 625 seconds. The labels are the same as in Fig. \ref{fig:sigev}.
\label{fig:mz}}
\end{figure}

In Figure \ref{fig:mz} we show the rate of the mass exchange between the disc
and corona, $\dot m_{\rm z}$, during the disc instability cycle. In the minimum
of the cycle the evaporation rate is very low and the maximal mass supply to the 
corona is achieved at $r \sim 6 R_{\rm Schw}$. When the disc outburst starts,
the evaporation rate grows dramatically in the middle of the unstable zone, while
dropping to  $\dot m_{\rm z} < 0$ at the outer edge of this zone. This is why the 
corona collapses locally at the outer parts, while expanding slightly in the inner 
parts during the disc outburst. In the decay phase the evaporation rate
again becomes low, with a decaying fluctuation at the outer edge
of the instability zone.

\subsection{Lightcurves}
\label{sec:lcurves}

 The lightcurves represent the luminosity of the disc and corona separately.
For the optically thick disc the luminosity is given by:
\begin{equation}
L_{\rm disc}=\int_{R_{\rm min}}^{R_{\rm max}} Q^{-}_{\rm rad} 2\pi r dr =
{4 \sigma \over 3 \kappa} \int_{R_{\rm min}}^{R_{\rm max}} {T^{4} \over \Sigma} 2\pi r dr
\label{eq:lumthin}
\end{equation} 

The luminosity of the corona is calculated under the assumption that the 
corona is in the thermal equilibrium. The ions are
 heated by the viscous dissipation and either cool by advection or transfer their
 energy to the electrons, which in turn radiate e.g. in the Inverse Compton process.
Therefore we have:
\begin{eqnarray}
 Q^{-}_{\rm cor} = Q^{+}_{\rm cor} - Q^{-}_{\rm adv}
 = {3 \over 2}\Omega \alpha_{\rm cor} P_{\rm cor} H_{\rm cor} - \\
\nonumber P_{\rm cor} r v_{\rm r}^{\rm cor} ({3 \over 2}
{\partial \ln T_{\rm cor} \over \partial r} - {\partial \ln \Sigma_{\rm cor} \over \partial r}) 
\end{eqnarray}
and
\begin{equation}
L_{\rm cor} = \int_{R_{\rm min}}^{R_{\rm max}} Q^{-}_{\rm cor} 2\pi r dr
\end{equation}

\begin{figure}
\epsfxsize = 250pt
\epsfbox{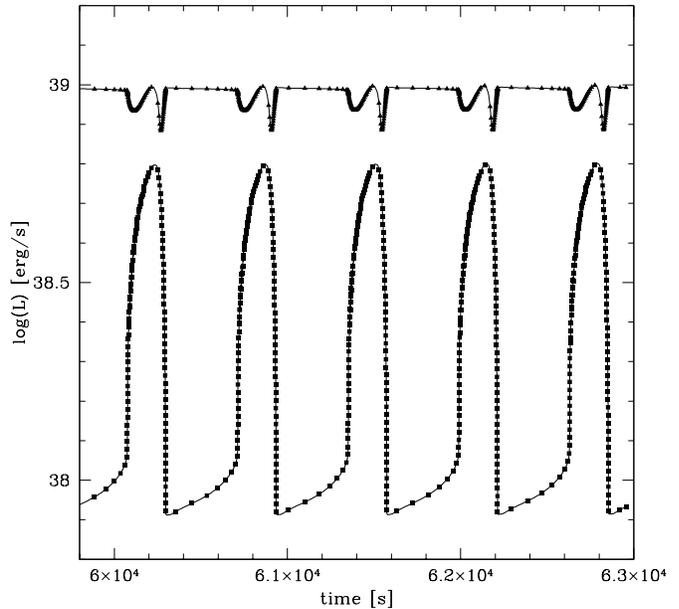}
\caption{The time evolution of the disc and corona luminosity. Lower curve shows 
the luminosity outbursts of the disc, while the upper curve shows the 
simultaneous coronal fluctuations.
 The external accretion rate is
$\dot M_{\rm ext}= 1.5 \times 10^{19}$ g/s.
\label{fig:lcurve}}
\end{figure}

In Figure  \ref{fig:lcurve} we show an exemplary lightcurve 
calculated  for several cycles of the disc outburst. The disc limit cycle 
is very strong since
$\dot M_{\rm ext}= 1.5 \times 10^{19}$ g s$^{-1}$ 
 substantially exceeds the critical value.
 Therefore the loops marked by the time-dependent solutions
on the $\Sigma-T$ plane encompass substantial range of temperatures and densities,
and the unstable region of the disc has the radial extension
up to about 80 $R_{\rm Schw}$.
This results in regular, large amplitude outbursts
of $\Delta \log L = 0.9$ and $\Delta t = 625$ s. 
The luminosity variations in the corona, 
however of the same frequency, are not that strong. Also, they are sometimes 
anti-correlated 
with the luminosity changes in the disc, since at the disc rise phase the coronal 
luminosity 
at first drops and then rises slightly, to drop and rise again during the disc decay 
phase.
Characteristically, between the disc outbursts the coronal luminosity 
decreases very slowly, with $\Delta \log L_{\rm cor}/\Delta t \approx 10^{-5}$,
whereas the
disc luminosity rises gradually, forming a 'wing' preceding the main outburst. 

\subsection{Accretion rate}
\label{sec:lcurves}

The mass exchange rate in our prescription depends on the
accretion rates in the disc and corona (Equation \ref{eq:dotmz}). Since they strongly
vary with time during the cycle of the disc evolution, the mass exchange can
have locally negative value. This has been shown in Figure \ref{fig:mz}.

\begin{figure}
\epsfxsize = 250pt
\epsfbox{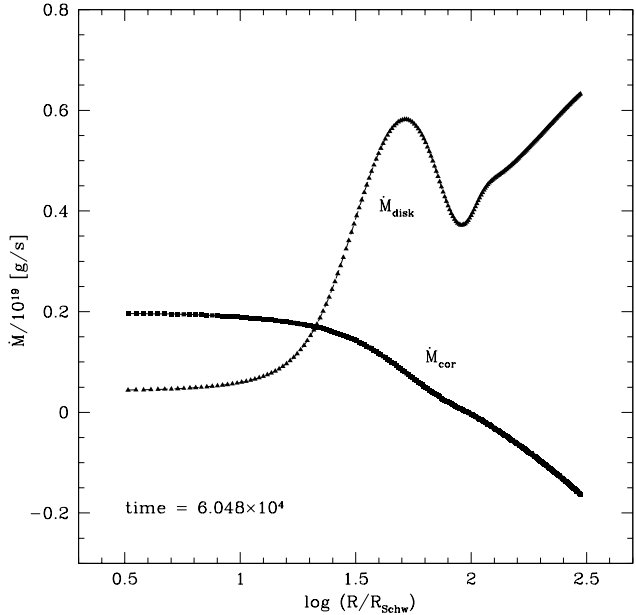}
\caption{Radial profiles of the accretion rate in the disc (triangles)
and in the corona (squares) between the disc outbursts. The time refers to
the lightcurve shown in Fig. \ref{fig:lcurve}.
\label{fig:mdotmin}}
\end{figure}

\begin{figure}
\epsfxsize = 250pt
\epsfbox{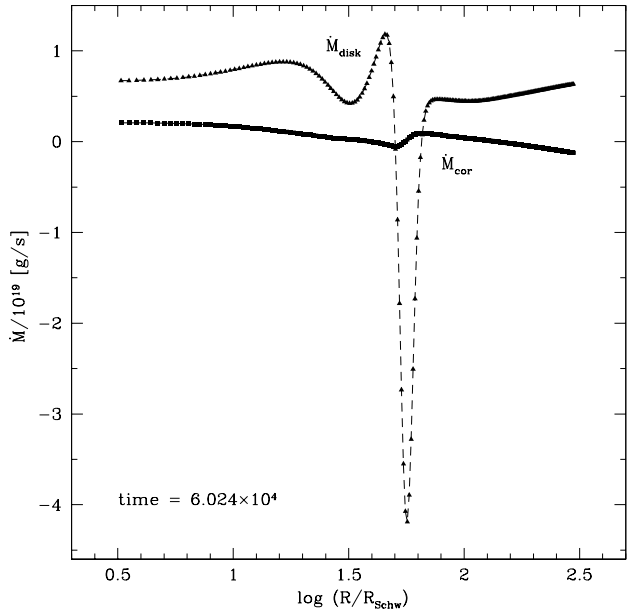}
\caption{Radial profiles of the the accretion rate in the disc (triangles)
and in the corona (squares) in the outburst. The time refers to
the lightcurve shown in Fig. \ref{fig:lcurve}.
\label{fig:mdotmax}}
\end{figure}

In Figures \ref{fig:mdotmin}  and \ref{fig:mdotmax} we show the radial profiles 
of the accretion rates in the disc and corona, taken in different phases of the cycle:
between the disc outbursts (corresponding to phase ``1'' in Fig.  \ref{fig:mz}) 
and in the 
outburst peak (phase ``3'').

In the minimum of the cycle, i.e. between the outbursts, the accretion rate
in the inner parts of the disc, where most of the dissipation takes place, is relatively low.
It rises substantially in the maximum of the cycle, therefore causing the disc luminosity
outburst.
The coronal accretion rate does not change that much, and the corresponding luminosity 
changes are not very pronounced.
This is because in certain radii, at the outer edge of the instability strip, the
radial velocity becomes negative. In consequence, the both accretion rates and
mass exchange rate, $\dot m_{z}$, also become negative, which means that 
some amount of material that has been evaporated to the corona, now goes back 
and sinks in the disc.
Negative values of $\dot m_{\rm z}$ are not unphysical: stationary models based on the
disc-corona mass exchange of R\'o\.za\'nska \& Czerny (2000) predict coronal 
condensation at some radii for high accretion rates.
The coronal surface density, and in turn the dissipation rate, is therefore
reduced, even in the inner regions, as the mass is spread out over the whole
corona. 
The temporary loss of the accretion rate can be estimated as 
$\Delta \dot M = 4\pi r \dot m_{\rm z} \Delta r$ and it can be as high as $1.9\times 
10^{18}$ g/s. Therefore almost the whole rise of the coronal accretion rate
(and luminosity), triggered by the disc outburst, is immediately compensated
by this loss due to the sinking of material in the strip around $\sim 50 R_{\rm Schw}$.

In addition, the accretion rate in the corona is negative in the outermost radii.
The extension of this zone of negative $\dot M_{\rm cor}$ depends on the
outer boundary condition in the corona, which in our case is fixed by 
$\Sigma_{\rm cor}(r_{\rm out}) = 1/\kappa \approx 2.94$. It implies that 
$\dot M_{\rm cor} < 0$ for $r > 100 R_{\rm Schw}$,
so the material slowly flows out from the corona at its outer radius.

\subsection{Alternative prescription for the mass exchange (prescription II)}
\label{sec:dissipat}

In the above used prescription for the rate of mass exchange between the disc 
and corona we
expressed the total locally generated flux as proportional to the sum of the 
accretion rates in the disc and corona. 
In case of a steady-state disc this is equivalent to the sum of the locally 
generated fluxes by the viscous energy dissipation. However, when we consider 
the time evolution of an 
unstable disc, the radial velocity, and in turn the accretion rate, can have 
locally negative  values in some parts of the disc. This forces the mass exchange 
rate to locally decrease 
in the state of the disc outburst, which is not the case for the viscously generated 
flux.
In other words, the prescription for the mass exchange rate in the outburst of an 
unstable 
disc is no longer equivalent to the sum of the locally dissipated fluxes.

Now we check whether the other, more 'conservative' prescription for the mass exchange 
rate can lead to different results of the time evolution of the disc plus corona 
system.
Instead the formula for the vertical mass transfer given by Equation \ref{eq:dotmz},
we use the following:
\begin{equation}
\dot m_{\rm z} = {3 \over 2} {\Omega r \over GM} f(r) 
\big(0.5 B_{1} \alpha_{\rm cor} \Sigma_{\rm cor} {GM \over r} + 
B_{2} \alpha_{\rm disc} P_{\rm disc} H_{\rm disc} \big)
\label{eq:dotmzalt}
\end{equation}
This form means that the evaporation is proportional to the sum of the energy 
dissipation rates in the disc and in the corona, and therefore is always positive.

\begin{figure}
\epsfxsize = 250pt
\epsfbox{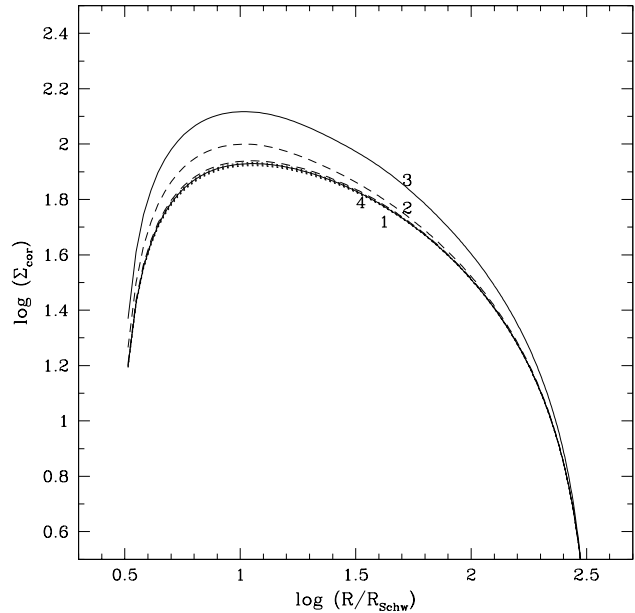}
\caption{Radial profiles of the corona surface density during the cycle of the 
evolution for the case of the mass exchange rate given by Eq. \ref{eq:dotmzalt}. 
The snapshots were made every 80 seconds and the whole cycle lasts
for 625 seconds. The labels are the same as in Fig. \ref{fig:sigev}.
\label{fig:sigmalt}}
\end{figure}

In Figure \ref{fig:sigmalt} we show the distribution of the coronal surface density  
during one cycle of the viscous-thermal instability (cf. Fig. \ref{fig:sigcorev}).
Clearly, there is no local dip in the corona that would be 
caused by the negative value of $\dot m_{\rm z}$
during the disc outburst, as it was in the previous case.
The material does not sink from the corona into the disc anywhere, but is uniformly evaporated
and the rate of the evaporation is the highest when the disc is the most luminous.

\begin{figure}
\epsfxsize = 250pt
\epsfbox{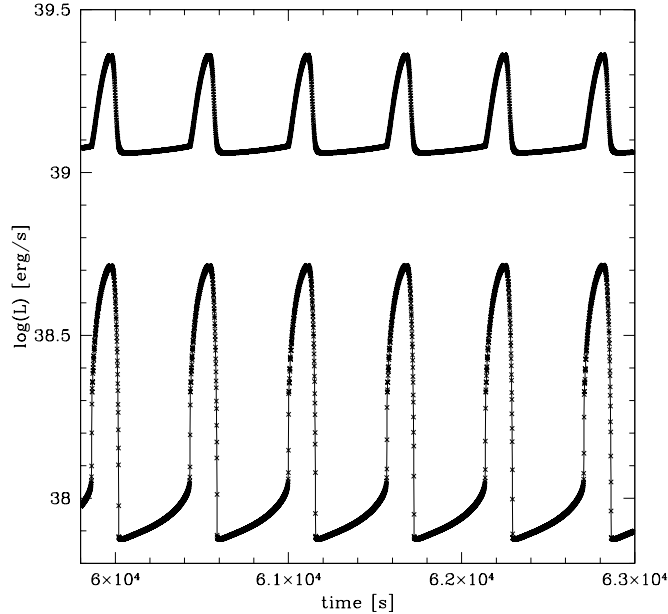}
\caption{The time evolution of the disc and corona luminosities,
for the case of the mass exchange rate given by Eq. \ref{eq:dotmzalt}.
 Lower curve shows 
the luminosity outbursts of the disc, while the upper curve shows the simultaneous
coronal outbursts.
 The external accretion rate is
$\dot M_{\rm ext}= 1.5 \times 10^{19}$ g/s.
\label{fig:lcurvealt}}
\end{figure}

Similarly, the total luminosity in the corona is now correlated with the disc luminosity.
In Figure \ref{fig:lcurvealt} we show the lightcurves analogous to those in Figure
\ref{fig:lcurve}. Between the outbursts the coronal luminosity is 
slightly increasing, and the lightcurve 
exhibits regular peaks, simultaneous to the disc outbursts.
The luminosity changes in the corona are of the order of $\Delta \log L_{\rm cor} = 0.3$.

\subsection{Hard versus soft luminosity: time lags and luminosity-color diagram}

Since the mass exchange prescription (II) determined by Equation (\ref{eq:dotmzalt})
gave more 
satisfactory representation of the time evolution 
with respect to the observed behaviour of GRS 1915+105, we adopt this prescription
 in further
analysis.

\begin{figure}
\epsfxsize = 250pt
\epsfbox{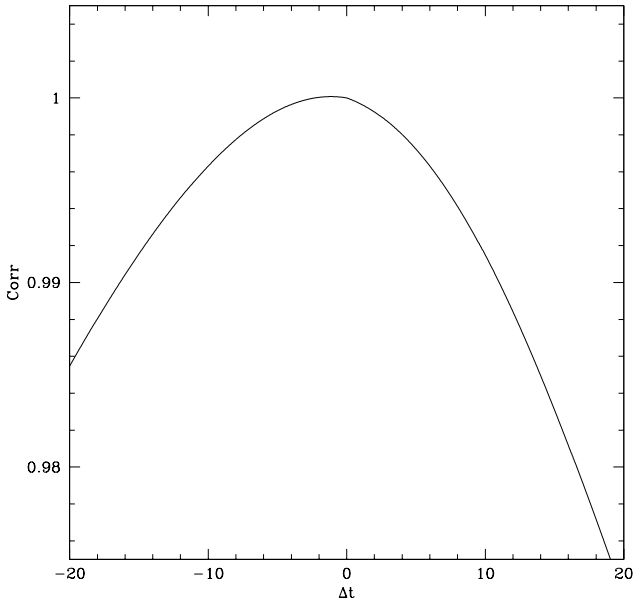}
\caption{The normalized cross-correlation function of the disc and corona lightcurves.
 The main peak is
shifted to -1.16 s, which means that the corona lightcurve lags the disc 
one by 1.16 s. 
\label{fig:corrmod}}
\end{figure}

We calculated the cross-correlation function, as defined by Equation
(\ref{eq:crossfun}), for the theoretical lightcurves
plotted in Figure \ref{fig:lcurvealt}. Our points were separated at least by 0.039 s,
and we found that the corona lightcurve lags the disc by 1.16 seconds. This is shown 
in Figure \ref{fig:corrmod}. The time lag in this case is equal to about 0.5\% of the duration 
of an outburst. We note that the outburst duration depends on the external accretion 
rate $\dot M_{\rm ext}$ and the lag depends on the coronal viscosity parameter 
$\alpha$, so we could obtain a range of values for different model parameters.

\begin{figure}
\epsfxsize = 250pt
\epsfbox{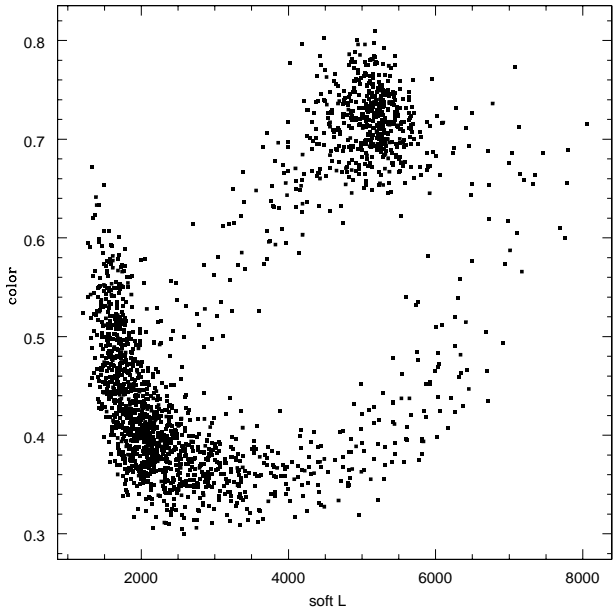}
\caption{The luminosity-color diagram of the two observed lightcurves,
shown in Fig. \ref{fig:obs}. 
On the horizontal axis is the soft X-ray flux in the range 1.5-6 keV,
and in the vertical axis is  the ratio of the hard to soft fluxes,
$F_{6.4 - 14.6 keV}/F_{1.5-6 keV}$.
\label{fig:color}}
\end{figure}

In Figure \ref{fig:color} we show the luminosity-color diagram for one of the 
observations  of GRS 1915+105, that shows the pronounced outbursts (class $\rho$, 
cf. Table \ref{tab:tab1}).
On the horizontal axis we plot the soft X-ray flux in the range 1.5-6 keV,
and on the vertical axis we plot  the X-ray color, i.e. the ratio of the hard 
to soft fluxes,
$F_{6.4 - 14.6 keV}/F_{1.5-6 keV}$.
During its evolution, the source follows a characteristic track, in the form of a loop
on this diagram. In the lower region (the ``banana'' shape) the soft luminosity is low,
and the X-ray color is soft. The upper region (``island'') shape is characterized 
by substantial luminosity both in hard and soft X-ray bands.

\begin{figure}
\epsfxsize = 250pt
\epsfbox{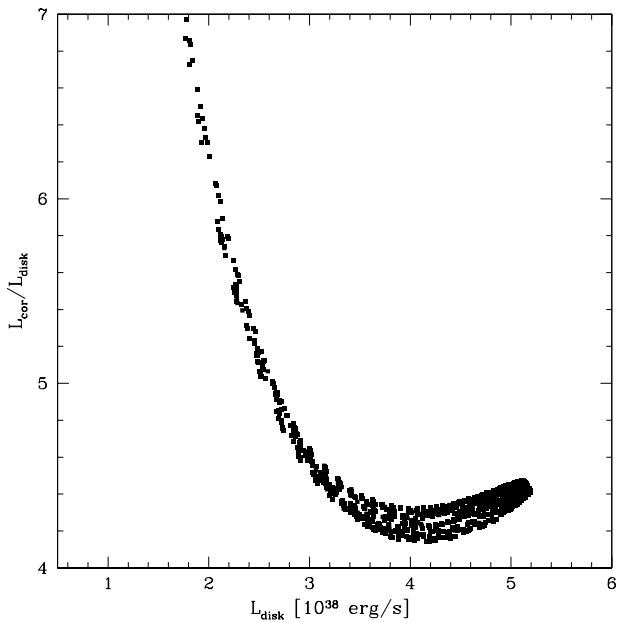}
\caption{The luminosity-color diagram of the two theoretical lightcurves, 
shown in Fig. \ref{fig:lcurvealt}. 
On the horizontal axis is the soft X-ray (disc) luminosity,
and in the vertical axis is  the ratio of the hard to soft (corona to disc) 
luminosities.
\label{fig:colormod}}
\end{figure}

In Figure \ref{fig:colormod} we show the result of our modeling. 
On the horizontal axis of the theoretical luminosity-color diagram we 
plot  the disc luminosity, while
in the vertical axis we plot the ratio of the corona to disc 
luminosities. The ``banana'' shape is reproduced in our model quite well.
However, we do not obtain the other distinct region on this diagram (the ``island'').
In our simulations the source moves only along  the track shown in Fig. 
\ref{fig:colormod}, and after reaching the end of the ``banana'' region
immediately jumps to the top-left corner of the diagram.

\section{Discussion}
\label{sec:diss}

The main simplification of our model was the assumption that the temperature in 
the corona 
is constant and equal to the virial one. Therefore the corona is heated only
via increase of its surface density and is not radiatively coupled to the disc.
We do not consider here any specific radiative processes that are important in 
cooling of the coronal gas, e.g. Comptonization.

On the other hand, this approach lets us solve the set of three time-dependent
equations for the disc plus corona evolution, accompanied by the formula for the mass 
exchange rate, and avoid numerical problems with the fourth equation for the
corona temperature. Using a finite time-step we are able to follow the time 
evolution of the 
system as long as it saturates in a quasi-stationary state, with oscillations
of constant amplitude and duration. We qualitatively and quantitatively check, 
how the behaviour of the underlying, unstable disc affects the corona and vice versa.

The full, two- or three dimensional treatment to the global long-term 
evolution of an accretion disc (or disc/corona system) is very complex.
Up to now, the sophisticated 2-D and 3-D accretion disc 
simulations (e.g. Agol et al. 2001;
Turner et al. 2003) either treated the problem locally, or did not include 
radiative cooling. Turner (2004) made the first attempt to compute
the 3-D simulation in  the flux-limited approximation; in this case however the 
plasma temperatures are still much too low to form the hot corona. 
Since the simulations are not able to reach the viscous timescale of the disc, the
long-term evolution of an X-ray source cannot be followed and any observational 
consequences of such a model are difficult to be checked.
Therefore the simple approach presented in this paper can still be valuable. 

\subsection{Viscosity parameterization}
\label{sec:viscpar}

In our choice of the viscosity law we follow the paper of Nayakshin,
Rappaport \& Melia (2000). The motivation of these authors was mostly
observational: they needed a viscosity law which allows for a disc 
instability at the intermediate accretion rates but provides the stable
solution at low and high accretion rates.
Moreover, the stable
upper branch should appear at accretion rates smaller or comparable to the
Eddington rate. The standard $\alpha P_{\rm tot}$ viscosity law of Shakura \& 
Sunyaev (1973) 
satisfies in a natural way all requirements but the last one. Upper stable 
branch in such solution forms due to advection 
(i.e. the ``slim disc''; Abramowicz et al. 1988), and the efficient advection
develops only for accretion rates much larger than the Eddington values,
definitely too high too account for the time behaviour e.g. of the microquasar
 GRS 1915+105.
What is more, in most cases the unstable branch should cease to exist at all
since most of the X-ray binaries accreting at high (but sub-Eddington) rates
are quite stable and well described by the classical disc (Gierli\' nski
\& Done 2004).

Therefore, observations tell us that some modifications are absolutely needed.
Either we must postulate very strong outflow (which effectively cools the 
disc), or a modification of the dissipation law itself.

We have much better understanding and
numerous observational constraints for the viscosity law when 
the gas pressure 
dominates. 3-D MHD simulations of the magneto-rotational instability (MRI)
well explain the nature
of the angular momentum transfer and the rough value of the viscosity 
parameter. Still, some ad hoc modifications, like
additional dependence on the disc thickness, are used to model the 
development of the ionization instability responsible for outbursts in 
cataclysmic variables and numerous X-ray transients (e.g. Cannizzo et al. 
1995).

The theoretical background of the viscosity law is still quite poor in case
when radiation pressure is important. Several authors in the past suggested 
that actually the whole idea of $\alpha P_{\rm tot}$  scaling is inappropriate,
and instead the $\alpha P_{\rm gas}$ should perhaps be used even if the radiation 
pressure dominates (Lightman \& Eardley 1974; Stella \& Rosner 1984). Sakimoto
\& Coroniti (1989) argued that the magnetic field will be expelled from the 
radiation pressure dominated disc by the buoyancy so the effectiveness of the
angular momentum transfer must decrease with an increase of radiation pressure.
Other authors argued that whenever radiation pressure dominates, new kinds of
instabilities develop which may modify the disc structure considerably.
Gammie (1998) discussed the photon bubble instability, and 
Ruszkowski \& Begelman (2003) argued this instability leads to the disc 
clumpiness which in turn 
influences the radiative transfer.

We expect some improvement of the viscosity modeling with the future 
development of MHD 
simulations of radiation pressure dominated discs. So far, two such simulations
were performed.  2-D computations of Agol et al. (2001) lasted only for a 
fraction of the
disc thermal timescale and as a result 
the disc collapsed to gas pressure dominated state.

A 3-D simulation of the disc which reached the thermal balance and lasted for
about 8 thermal timescales was recently 
completed by Turner (2004). The disc did not achieve
the full stability and showed long-lasting variations by a factor of a few 
but the instability was not as violent as predicted by the standard 
$\alpha P_{\rm tot}$ mechanism (Szuszkiewicz \& Miller 1998; Janiuk et al. 2002). 
The magnetic field lines were indeed partially
expelled from the disc interior. In this simulation, the initial 
Shakura \& Sunyaev (1973) state
with $\alpha = 0.01$ evolved to a complex state with mean dissipation level 
equivalent to $\alpha = 0.0013$. 

Therefore, the  modified viscosity law, given by Eq.~\ref{eq:visclaw}, 
seems to be a plausible option in case of the accretion disc.
Since in the corona the radiation pressure does not contribute to the total pressure,
the classical viscosity parameterization of the corona 
with constant $\alpha$ is also justified.

\subsection{Comparison with observations}

Our time dependent model of the non-stationary accretion onto a black hole
gives at least a partial explanation of the complex variability
of the microquasar GRS 1915+105. Direct evidence for variable accretion rate in 
this microquasar comes from the
spectral analysis (Migliari \& Belloni 2003), and a plausible 
mechanism that leads to the local accretion rate variations in the inner parts of a 
disc, giving the outbursts of appropriate amplitudes and durations,
is the radiation pressure instability.

The microquasar is unique because of its
large amplitude, regular outbursts. Such a behaviour is not observed in other 
Galactic X-ray sources, probably because the Eddington ratio in 
this microquasar  is higher than in other  black hole systems (Done, Wardzi\' nski 
\& Gierli\' nski 2004).
Our model, at least qualitatively, explains this phenomenon.
The radiation pressure instability leads to the strong outbursts only 
if the accretion rate is higher than some critical value. Also,
the corona has a stabilizing role and can suppress the disc oscillations completely
or make them less pronounced.
Quantitatively, however, the reliable determination of the critical accretion rate
above which the outbursts occur, would require a more detailed modeling
of viscosity within the disc and essentially the 2-dimensional calculations.
This is beyond the scope of the present work.


The physical coupling between the disc and corona via the mass exchange leads
in a natural way to the regular time delays between the disc (soft X-ray)
and corona (hard X-ray) emission. The lag of the order of 1 s is 
comparable with the viscous timescale in the corona and is required by this
hot flow to adjust to the variable conditions in the underlying disc.
Such lags are present in some observed lightcurves of the microquasar, i.e.
these exhibiting outbursts that are possibly connected with the radiation pressure
instability.

The strong variability of GRS 1915+105 manifests itself also
by characteristic tracks on the color-color and luminosity-color diagram.
The bottom-left corner of the luminosity-color plot is occupied by the source
during its gradual rise (the ``wing''), preceding the outburst state.
This, so called here ``banana'' shape is well modeled in our calculations.
While the disc luminosity (soft X-ray flux) is gradually rising before the outburst,
the rise in the coronal luminosity is much flatter, leading therefore to the decreasing
hard-to-soft flux ratio (color). 
On the other hand, in the outburst peak, when the soft luminosity is the
highest, the color starts rising. Just after the outburst there appears 
a somewhat flat maximum in both hard and soft X-rays, leading to
the other distinct region occupied by the source on the luminosity color-diagram:
so called here ``island'' shape. This is not present in our model calculations, since
our outburst maximum is a sharp peak. Therefore we do not obtain a state of both
high disc and corona fluxes, which in consequence would give the high color in the 
disc emission peak. Instead, the corona lags the disc emission and its maximum
corresponds to the already decaying state of the disc.

\section{Conclusions}
\label{sec:con}

We presented the first results of the time-dependent calculations of thermal-viscous
evolution of accretion disc that is coupled to the corona by the mass exchange.
Contrary to the accretion disc, the corona is stable, since there is only gas 
pressure included in 
its equation of state. However, it is also 
to a certain extent subject to similar changes as the accretion disc
during its instability cycle, since it actively responds to the behaviour of 
the underlying disc.

The main conclusions of the present work are as follows:
\begin{itemize}
\item The luminosity outbursts in the disc are correlated with similar outbursts in the corona
if the mass exchange rate is proportional to the sum of the locally dissipated fluxes
( prescription II).
The assumption that it is proportional to the sum of the local accretion rates 
(prescription I) is equivalent, 
but only during the disc quiescence. In the outburst this prescription leads
to the luminosity dips in the corona rather than outbursts.
Specifically, the lightcurves of the disc and corona are anti-correlated,
with two dips in the corona corresponding to the peak of the disc luminosity.
\item The luminosity profile between the outbursts in the corona is 
gradually decreasing,
while in 
the disc it is gradually rising (prescription I). 
 Alternatively, the prescription II leads to the slow rise of the coronal flux,
correlated with the steep rise of the disc.
\item The coronal outbursts are of much smaller amplitude than the disc ones, but
of similar duration.
\item The outbursts in the corona lag the disc evolution by $\sim 1$ second,
which is in good agreement with observations of the microquasar GRS 1915+105.
\item The outer boundary condition in the corona determines if the material flows out
at the outer edge. For small surface density $\Sigma_{\rm out} \sim 3$ g/cm$^{2}$
 the accretion rate in the corona is negative above $\sim 100 R_{\rm Schw}$.
\item If the mass exchange between the disc and corona is proportional to the sum of 
the 
local accretion rates  (prescription I), the material may also locally sink into the 
disc in the region
around $50 R_{\rm Schw}$.
\end{itemize}

Our model is only in part able to reproduce the tracks of the system on the 
luminosity-color diagram,
having a ``banana'' shape without the ``island'' one. This should be modeled in more 
detail, including also the spectral analysis.
In addition, the tracks on the color-color diagram, not analyzed here, 
in some observed sources may have a form of {\it hysteresis} 
(Maccarone \& Coppi 3002; Zdziarski et al. 2004).
Since the phenomenon is connected with much longer timescales, than those considered here,
the underlying instability mechanism should be different. A plausible one
seems to be the partial hydrogen ionization instability, which operates essentially in the same
way, causing the disc to oscillate between the cold and hot phases.
This is the subject of our future investigations.

\section*{Acknowledgments}
We thank  Ma{\l}gosia Sobolewska for help in data reduction and 
Agata R\'o\.za\'nska, Piotr \.Zycki, Friedrich Meyer and Emmi Meyer-Hofmeister
for helpful discussions. 
This research has made use of 
data obtained through the HEASARC Online Service, provided by NASA/Goddard 
Space Flight Centre.
This work was supported in part by  
grant 2P03D00124  of the Polish State Committee for Scientific Research.


\begin{thebibliography}{99}

\bibitem[\protect\citeauthoryear{Abramowicz et al.}{1988}]{a1} Abramowicz M. A., 
Czerny B., Lasota J.-P., Szuszkiewicz E., 1988, ApJ, 332, 646
\bibitem[\protect\citeauthoryear{Agol et al.}{2001}]{a2} Agol E., Krolik J., Turner N.J., Stone J.M., 2001, ApJ, 558, 543
\bibitem[\protect\citeauthoryear{Belloni et al.}{2000}]{b1} Belloni T., Klein-Wolt M., Mendez M., van der Klis M., van Paradijs J., 2000, A\&A, 355, 271 
\bibitem[\protect\citeauthoryear{Cannizzo et al.}{1995}]{c1} Cannizzo J.K., Chen W., Livio M., 1995, ApJ, 454, 880
\bibitem[\protect\citeauthoryear{Chen \& Taam}{1993}]{c2} Chen X., Taam R.E., 1994, 
ApJ, 412, 254
\bibitem[\protect\citeauthoryear{Done}{2002}]{d1} Done C., 2002, in `` X-ray astronomy in the new millennium'', Eds. R. D. Blandford, A. C. Fabian and K. Pounds, Roy. Soc. of London Phil. Tr. A., vol. 360, Issue 1798, p.1967
\bibitem[\protect\citeauthoryear{Done}{2004}]{d2} Done C., Wardzi\'nski G., Gierli\'nski M., 2004, MNRAS, 349, 393
\bibitem[\protect\citeauthoryear{Gammie}{1998}]{g1}Gammie C.F.,  1998, MNRAS, 297, 929
\bibitem[\protect\citeauthoryear{Gierli\' nski \& Done}{2004}]{g2}Gierli\' nski M., Done C., 2004, MNRAS, 347, 885
\bibitem[]{H3} Honma F., Matsumoto R., Kato S. 1991, PASJ, 43, 147
\bibitem[\protect\citeauthoryear{Janiuk \& Czerny}{2000}]{j1} Janiuk A., Czerny B.,
2000, New Astronomy, 5, 7
\bibitem[\protect\citeauthoryear{Janiuk et al.}{2002}]{j2} Janiuk A., Czerny B., Siemiginowska A., 2002, ApJ, 576, 908
\bibitem[\protect\citeauthoryear{Janiuk et al.}{2004}]{j3} Janiuk A., Czerny B., Siemiginowska A., Szczerba R., 2004, ApJ, 602, 595
\bibitem[]{M2} Meyer F., Meyer-Hofmeister E., 1984, A\&A, 132, 143
\bibitem[\protect\citeauthoryear{Maccarone \& Coppi}{2003}]{mac} Maccarone T.J., 
Coppi P.S., 2003, MNRAS, 338, 189 
\bibitem[\protect\citeauthoryear{Meyer \& Meyer-Hofmeister}{1994}]{m1} Meyer F.,
Meyer-Hofmeister E., 1994, A\&A, 288, 175
\bibitem[\protect\citeauthoryear{Mirabel \& Rodriguez}{1994}]{m2} Mirabel I.F., Rodriguez L.F., 1994, Nature, 371, 46
\bibitem[\protect\citeauthoryear{Migliari}{2003}]{mig} Migliari S., Belloni T., 2003, A\&A, 404, 283
\bibitem[\protect\citeauthoryear{Muchotrzeb \& Paczy\' nski}{1982}]{m3}  Muchotrzeb B., Paczy\' nski B., 1982, Acta Astron., 32, 1
\bibitem[\protect\citeauthoryear{Muno et al.}{2001}]{m4} Muno M., et al., 2001, ApJ, 556, 515 
\bibitem[\protect\citeauthoryear{Naik et al.}{2002}]{n2} Naik S., Agrawal P.C., Rao A.R., Paul B., 2002, MNRAS, 330, 487
\bibitem[\protect\citeauthoryear{Nayakshin et al.}{2000}]{n1} Nayakshin S., Rappaport 
S.,  Melia F., 2000, ApJ, 535, 798
\bibitem[\protect\citeauthoryear{Lasota \& Pelat}{1991}]{las} Lasota J.-P., Pelat D.,
1991, A\&A, 249, 574
\bibitem[\protect\citeauthoryear{ Lightman \& Eardley}{1974}]{l3} Lightman A.P., Eardley D.M., 1974, ApJ, 187, L1
\bibitem[\protect\citeauthoryear{Paczy\' nski \& Bisnovatyi-Kogan}{1981}]{p0} Paczy\' nski B. \& Bisnovatyi-Kogan G., 1981, Acta Astron., 31, 283
\bibitem[\protect\citeauthoryear{Paczy\' nski \& Wiita}{1980}]{p1}  Paczy\' nski B., 
Wiita P.J., 1980, A\&A, 88, 23
\bibitem[\protect\citeauthoryear{Pringle, Rees \& Pacholczyk}{1974}]{p6} Pringle J.E., Rees M.J., Pacholczyk, A.G., 1974, A.\&A., 29, 179
\bibitem[\protect\citeauthoryear{R\'o\.za\'nska \& Czerny}{2000}]{r1}R\'o\.za\'nska A., Czerny B., 2000, A\&A, 360, 1170
\bibitem[\protect\citeauthoryear{Ruszkowski \& Begelman}{2003}]{r2} Ruszkowski M., Begelman M.C., 2003, ApJ, 586, 384
\bibitem[\protect\citeauthoryear{Sakimoto \& Coroniti}{1989}]{s0} Sakimoto P.J., Coroniti F.V., 1989, ApJ, 342, 49
\bibitem[\protect\citeauthoryear{Shakura \& Sunyaev}{1973}]{s1} Shakura N.I., Sunyaev R.A., 1973, A.\&A., 24, 337
\bibitem[]{S3} Shakura N.I., Sunyaev R.A., 1976, MNRAS, 175, 613
\bibitem[]{S8} Smak J. 1984, Acta Astron. 34, 161
\bibitem[]{Sob} Sobolewska M., \.Zycki P.T., 2003, A\&A, 400, 553
\bibitem[\protect\citeauthoryear{Stella \& Rosner}{1984}]{s2} Stella L., Rosner R., 1984, ApJ, 277, 312
\bibitem[\protect\citeauthoryear{Stull}{1988}]{stull} Stull R.B., 1988,
{\it An Introduction to Boundary Layer Meteorology}, Kluwer Academic Publishers, 
Dordrecht
\bibitem[\protect\citeauthoryear{Szuszkiewicz\& Miller}{1998}]{s3} Szuszkiewicz E., Miller J., 1998, MNRAS, 298, 888
\bibitem[\protect\citeauthoryear{Taam \& Lin}{1984}]{taam1} Taam R. E.,  Lin D. N. C.,
 1984, ApJ, 287, 761
\bibitem[\protect\citeauthoryear{Taam et al.}{1997}]{taam2} Taam R. E., Chen X., 
Swank J.H., 1997, ApJ, 485, L83
\bibitem[\protect\citeauthoryear{Turner et al.}{2003}]{t0}Turner N.J., Stone J.M., 
Krolik J.H., Sano T., 2003, ApJ, 593, 992
\bibitem[\protect\citeauthoryear{Turner}{2004}]{t1} Turner N.J., 2004, ApJ, 605, L45
\bibitem[\protect\citeauthoryear{Watarai}{2003}]{w1} Watarai \& Mineshige S., 2003, ApJ, 596, 421
\bibitem[\protect\citeauthoryear{Zdziarski et al.}{2004}] {z1} Zdziarski A.A., 
Gierli\'nski M., Miko{\l}ajewska J., Wardzi\'nski G., Harmon A.B., Kitamoto S., 2004,
MNRAS, 351, 791
\end{thebibliography}
\end{document}